\documentclass[aps,prx,reprint,superscriptaddress,longbibliography,floatfix]{revtex4-2}

\newcommand{\ket}[1]{\left \lvert#1 \right \rangle}

\newcommand{\bbraket}[3]{\langle#1 | #2 | #3 \rangle}

\newcommand{\abs}[1]{\left \lvert#1 \right \rvert}

\newcommand{\qzpf}{Q_\text{zpf}}
\newcommand{\phizpf}{\Phi_\text{zpf}}

\makeatletter
\newenvironment{figurehere}
{\def\@captype{figure}}
{}
\makeatletter

\DeclareRobustCommand{\quickfig}[4]{\begin{figure}
\begin{centering}
\includegraphics[width=#1]{#2}
\par\end{centering}
\caption{#3}\label{#4}
\end{figure}
}

\DeclareRobustCommand{\quickwidefig}[4]{\begin{figure*}[ht]
\begin{centering}
\includegraphics[width=#1]{#2}
\par\end{centering}
\caption{#3}\label{#4}
\end{figure*}
}

\usepackage{amsmath}
\usepackage{amstext}
\usepackage{amssymb}
\usepackage{appendix}
\usepackage{coseoul}
\usepackage{graphicx}
\usepackage{import}
\usepackage{modular}

\usepackage[pdfpagemode=UseNone,pdfstartview=FitH,colorlinks=true,linkcolor=blue,citecolor=blue,urlcolor=blue]{hyperref}
\usepackage[all]{hypcap}

\begin{document}
\title{Balanced Coupling in Electromagnetic Circuits}
\author{Daniel Sank}
\author{Mostafa Khezri}
\author{Sergei Isakov}
\author{Juan Atalaya}
\affiliation{Google Quantum AI, Goleta, CA}
\date{\today}

\begin{abstract}
The rotating wave approximation (RWA) is ubiquitous in the analysis of driven and coupled resonators.
However, the limitations of the RWA seem to be poorly understood and in some cases the RWA disposes of essential physics.
We investigate the RWA in the context of electrical resonant circuits.
Using a classical Hamiltonian approach, we find that by balancing electrical and magnetic components of the resonator drive or resonator-resonator coupling, the RWA can be made exact.
This type of balance, in which the RWA is exact, has applications in superconducting qubits where it suppresses nutation normally associated with strong Rabi driving.
In the context of dispersive readout, balancing the qubit-resonator coupling changes the qubit leakage induced by the resonator drive (MIST), but does not remove it in the case of the transmon qubit.
\end{abstract}
\maketitle

\levelstay{Introduction}

Analysis of driven and coupled resonators is pervasive throughout physics and engineering, and the case of electrical resonators has attracted particularly strong attention due to the success of superconducting circuits in quantum computing \cite{Krantz:review:2019, Blais:review:2020}.
Theoretical analyses almost universally use the rotating wave approximation (RWA) wherein fast terms in the system's equations of motion are neglected.
The RWA simplifies the equations of motion for driven and coupled resonators, and for coupled resonators it greatly simplifies expressions for the normal mode frequencies.
Nevertheless, the validity of the RWA seems to be poorly understood.
The RWA is invoked in the coupled mode theory at the foundation of design of parametrically driven and non-reciprocal devices, used for example in Refs.~\cite{Lecocq:non_reciprocal:2017, Peterson:non_reciprocity:2017, Ranzani:graph:2015}.
Each of these references ultimately refers to Ref.~\cite{Louisell:coupled_modes} which justifies the RWA entirely by the fact that the RWA leads to a good approximation for the normal mode frequencies of coupled resonators.

But how well does the RWA capture the dynamics of real devices?
The RWA is known to fail in the case of superconducting qubit readout, where a weakly anharmonic resonator (transmon qubit) is coupled to a harmonic resonator.
In that case, higher order transitions between quantum levels in the coupled qubit-resonator system were found to spoil the readout process \cite{Sank:rotating_wave:2016} even though those transitions are absent in the traditional RWA analysis \cite{Schuster:ac_Stark_dephasing:2005, Wallraff:coupling:2004, Blais:cQED:2004, Khezri:dressed_squeezed_states:2016}.
The RWA has also come under some scrutiny recently \cite{Zeuch:2020:rotating_wave} as lowering the frequency of superconducting qubits has been pursued as a way to increase their energy decay lifetimes.
As the qubit frequency approaches the frequency of driven Rabi oscillations used for logic gates, the assumptions of the RWA are violated and it cannot be expected to be accurate.
With these examples in mind, a natural question is whether we can engineer the driving or coupling of electromagnetic resonators to actually remove the fast terms at the hardware level, i.e. to make the RWA exact.
If we can do this, then not only would the devices be easier to analyze, but also the deleterious processes found in Ref.~\cite{Sank:rotating_wave:2016} might be suppressed.

In this paper, we show that by balancing the electric and magnetic aspects of driven and coupled resonators, we can remove the fast oscillating terms and make the RWA exact.
The paper is organized as follows.
Section \ref{sec:driving} provides mathematical analysis of a driven resonator, showing how proper balance between the amplitudes and phases of simultaneous capacitive (electric) and inductive (magnetic) drives makes the RWA exact.
Section \ref{sec:coupling} analyzes two coupled resonators, showing that the RWA can be made exact if the coupling uses a balance of capacitive and inductive parts.
Application to readout of superconducting qubits is discussed.
Section \ref{sec:coupled_lines} shows that the RWA appears in the analysis of coupled transmission lines, and that the design of a directional coupler is equivalent to making the RWA exact.
Finally, section \ref{sec:conclusions} offers conclusions.

\levelstay{Balanced Drive} \label{sec:driving}

\quickfig{\columnwidth}{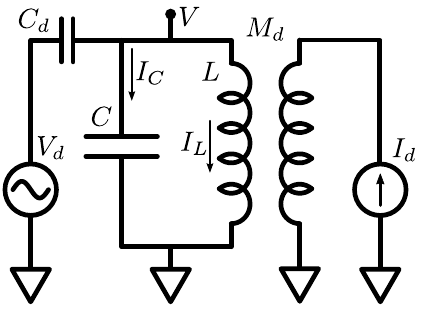}{Capacitively and inductively driven resonator.}{fig:driven_resonator}

An LC resonator driven both capacitively and inductively is shown in Figure \ref{fig:driven_resonator}.
Kirchhoff's laws for this circuit are
\begin{align}
  C_d \left( \ddot{V}_d - \ddot{V} \right) &= \dot{I}_C + \dot{I}_L \\
  L \dot{I}_L + M_d \dot{I}_d &= V \, .
\end{align}
Going through the usual process of guessing the Lagrangian and then converting to the Hamiltonian, we find
\begin{align}
  H
  &= \underbrace{\frac{Q^2}{2C'} + \frac{\Phi^2}{2L}}_{H_0} \nonumber \\
  &+ \underbrace{\frac{C_d}{C'}V_d(t) Q
  - \frac{M_d}{L} I_d(t) \Phi}_{H_d} \label{eq:drive_hamiltonian_lab_frame}
\end{align}
where $C' \equiv C + C_d$.
Details are given in Appendix \ref{appendix:driving}.
The symmetry between the flux and charge terms, and between the voltage and current drive terms, is brought out by defining the impedance $Z' \equiv \sqrt{L / C'}$, flux and charge scales $\phizpf \equiv \sqrt{\hbar Z' / 2}$ and $\qzpf \equiv \sqrt{\hbar / 2 Z'}$ \footnote{$\Phi_\text{zpf}$ and $Q_\text{zpf}$ are the flux and charge zero point fluctuations in the ground state of the quantum LC resonator.}, and dimensionless variables $X \equiv \Phi / 2 \phizpf$ and $Y \equiv Q / 2 \qzpf$.
Using these variables, the Hamiltonian is
\begin{equation}
  H = \underbrace{\hbar \omega_0 (X^2 + Y^2)}_{H_0}
  + \underbrace{2 \, G_y y(t) Y - 2 G_x \, x(t) X}_{H_d}
\end{equation}
where
\begin{align*}
  G_x & \equiv (M_d / L) \phizpf I_d \\
  G_y & \equiv (C_d / C') \qzpf V_d
\end{align*}
and
\begin{align*}
  I_d(t) & = I_d \, x(t) \\
  V_d(t) & = V_d \, y(t)
\end{align*}
renormalize the drive magnitudes.
Here $V_d$ and $I_d$ are arbitrary constants with dimensions of voltage and current introduced so that $x(t)$ and $y(t)$ are dimensionless.
The equations of motion are
\begin{equation}
  \dot{X} =  \frac{1}{2 \hbar} \frac{\partial H}{\partial Y}
  \qquad
  \dot{Y} = -\frac{1}{2 \hbar} \frac{\partial H}{\partial X}
  \, ,
\end{equation}
and in the absence of the drive the solutions are $X(t) =   X(0) \cos(\omega_0 t)$ and $Y(t) = - Y(0) \sin(\omega_0 t)$ where $\omega_0 = 1 / \sqrt{L C'}$.
This motion is analogous to a spin precessing in space as described in Appendix~\ref{appendix:overview}, and just as is done in analysis of a driven precessing spin, it's convenient here to use a coordinate frame rotating with the intrinsic motion.
Passage to the rotating coordinate frame is easier with the mode variables $a$ and $a^*$ defined by 
\begin{align}
    X = (a + a^*)/2, \;\;\; Y = -i(a - a^*)/2. \nonumber 
\end{align}
With these variables, the Hamiltonian is
\begin{equation}
  H = \underbrace{\hbar \omega_0 \abs{a}^2}_{H_0}
  \underbrace{- z(t) a - z(t)^* a^*}_{H_d}
\end{equation}
where $z(t) \equiv G_x \, x(t) + i G_y \, y(t)$.
The equations of motion are
\begin{equation}
  \dot{a} = -\frac{i}{\hbar} \frac{\partial H}{\partial a^*}
  \qquad
  \dot{a}^* = \frac{i}{\hbar} \frac{\partial H}{\partial a}
\end{equation}
and in the absence of the drive the solution is $a(t) = a(0) \exp(-i \omega_0 t)$.
We switch to the rotating frame by defining $\bar{a}(t) \equiv a(t) \exp(i \omega_r t)$ and the Hamiltonian in the rotating frame is
\begin{equation}
  H_r
  =
  \underbrace{\hbar(\omega_0 - \omega_r) \abs{\bar{a}}^2}_{H_{r,0}}
  \underbrace{- z(t) \bar{a} e^{-i \omega_r t} - z(t)^* \bar{a}^* e^{i \omega_r t}}_{H_{r,d}} \, .
  \label{eq:hamiltonian_oscillator_driven_rotating_frame}
\end{equation}
In typical experiments either the electric or the magnetic part of the drive is present, but not both.
For example, the electric part may be absent, i.e. $V_d(t) = 0$, and the magnetic part may be sinusoidal, $I_d(t) = I_d \cos(\omega_d t)$.
In this case
\begin{align}
  H_{r,d}
  =& - G_x \cos(\omega_d t) e^{-i \omega_r t} \bar{a} + \text{c.c.} \nonumber \\
  =& - (G_x / 2)
  \left(
    e^{-i(\omega_r - \omega_d)t} \bar{a} + e^{i(\omega_r - \omega_d)t} \bar{a}^*
  \right. \nonumber \\
  & \left.
   + e^{-i(\omega_r + \omega_d)t} \bar{a} + e^{i(\omega_r + \omega_d)t} \bar{a}^*
  \right)
  \, .
\end{align}
Superconducting quantum devices have resonance frequencies in the $\sim 5 \, \text{GHz}$ range and control pulses are typically applied close to resonance, so while the slow terms may have frequencies $(\omega_r - \omega_d)/2\pi \lesssim 10\,\text{MHz}$, the fast terms oscillate at $(\omega_r + \omega_d)/2\pi \sim 10 \text{GHz}$.
The RWA drops the fast-oscillating terms, typically with the justification that, in the weak driving limit where $G_x \ll \hbar \omega_0$, such fast terms cannot significantly affect the dynamics.
It should be noted, however, that recent interest in lowering the resonance frequency of quantum devices to below $1 \, \text{GHz}$ while maintaining strong driving \cite{Zhang:fluxonium:2021, Najera:fluxonium:2024, Connolly:readout:2024, Campbell:2020:nonadiabatic} pushes the limits of the weak driving assumption and so non-RWA effects may come into play.

However, if both the electric and magnetic parts of the drive are included with equal magnitude and one quarter cycle phase offset, i.e.
\begin{align}
  G_x &= G_y \equiv G_d \nonumber \\
  x(t) &= \cos(\omega_d t + \phi_d) \nonumber \\
  y(t) &= \sin(\omega_d t + \phi_d)
  \, , \label{eq:drive_condition}
\end{align}
then $z(t) = G_d \exp(i (\omega_d t + \phi_d))$ and the drive Hamiltonian is
\begin{equation}
  H_{r,d} = - G_d \left( e^{-i ( \omega_r - \omega_d)t} e^{i \phi_d} \bar{a} + \text{c.c.} \right)
\end{equation}
where the fast terms have vanished exactly.
Condition (\ref{eq:drive_condition}) says that for the RWA to be exact, the electric and magnetic drive strengths must be equal, and the charge drive must lag the flux drive by one quarter of a cycle ($\pi/2$ radians), a condition we call ``balanced drive''.
This is the first main result of the paper: true rotating drive fields can be constructed in electromagnetic resonators by balancing the strengths of the electric and magnetic parts of the drive.
The result applies to any one dimensional system so long as the drive can be coupled to both the position and its conjugate momentum, but the electromagnetic case is particularly interesting because simultaneous capacitive and inductive drive coupling is routine in experiments.

For an unbalanced purely inductive drive turning on at $t=0$, i.e. $z(t) = G \cos(\omega_0 t)$ for $t \geq 0$ and $z(t) = 0$ for $t < 0$, the time evolution of the resonator amplitude is
\begin{equation}
  a(t) = \frac{i G}{2} \left( t + \frac{e^{i 2 \omega_0 t} - 1}{i 2 \omega_0} \right)
  \, .
  \label{eq:dynamics_solution}
\end{equation}
Invoking the RWA or using balanced drive with $z(t) = (G/2) \exp(-i \omega_0 t)$ drops the second term.
The dynamics with $\omega_0 = (2\pi)\times 1$ and $G = (2\pi)\times 0.1$ are shown in Fig.~\ref{fig:dynamics}.
When the drive is balanced or when we make the RWA with an unbalanced drive, the oscillator's imaginary part increases linearly while the real part remains zero (solid blue curve).
On the other hand, when the drive is not balanced and we do not invoke the RWA, the real and imaginary parts both suffer an oscillation with period $1 / 2 \omega_0$ (broken orange curve).

\quickfig{\columnwidth}{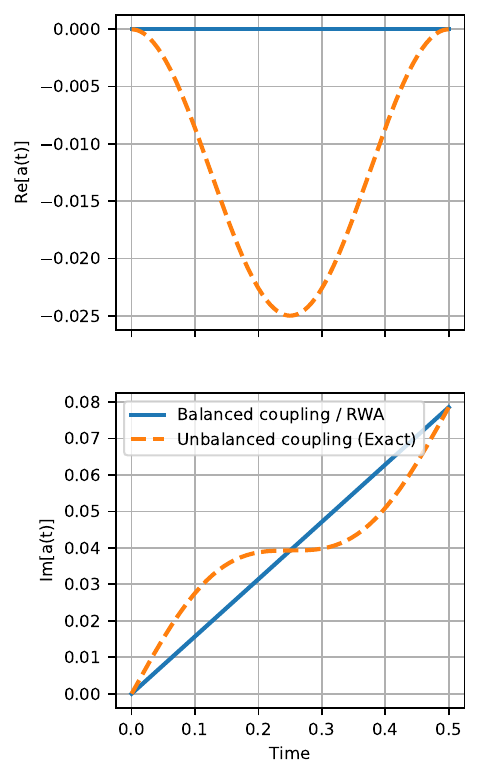}{Dynamics of a driven harmonic resonator. The blue solid curve corresponds to balanced drive or invoking the RWA with unbalanced drive. The orange broken curve shows the exact solution with unbalanced drive, i.e. Eq.~(\ref{eq:dynamics_solution}).}{fig:dynamics}

Balanced drive may allow strong driving of low frequency superconducting circuits such as fluxonium, or even transmons operated at low frequency to enhance their energy decay lifetimes $T_1$.
The Hamiltonian for the driven two level system may be obtained from Eq.~(\ref{eq:hamiltonian_oscillator_driven_rotating_frame}) through the replacements $\bar{a} \rightarrow \sigma_-$ and $\bar{a}^* \rightarrow \sigma_+$.
Figure \ref{fig:dynamics_tls} shows the expectation values of the Pauli operators for a two level system driven on resonance with $\omega_0 = (2\pi) \times 1$ and $G = (2\pi) \times 0.1$.
In the exact solution with unbalanced drive (blue, green, black) the trajectory exhibits nutation, i.e. each component suffers a fast oscillation on top of the usual rotation around the Bloch sphere.
On the other hand, with the balanced drive (red) the nutation is removed and the trajectories are smooth.
These predictions were recently verified in an experiment with fluxonium, where the balanced drive was shown to remove nutation in Rabi oscillations \cite{Rower:counterrotating:2024}.
\quickfig{\columnwidth}{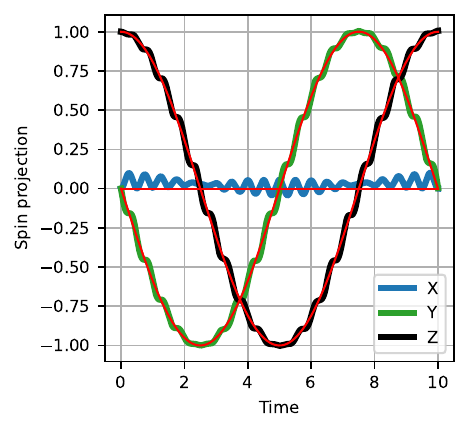}{Trajectory of a two level system driven on resonance. The blue, green, and black curves show exact solutions of the $X$, $Y$, and $Z$ projections of the system in the case of unbalanced drive. The thin red curves show the trajectories when invoking the RWA or when using balanced drive.}{fig:dynamics_tls}

\levelstay{Two-mode balanced coupling} \label{sec:coupling}

\quickfig{\columnwidth}{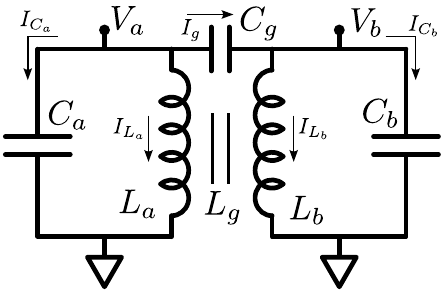}{Two resonators coupled capacitively and inductively.}{fig:coupledOscillatorsLAndC:diagram}

We now consider the case of two coupled LC resonators as shown in Figure \ref{fig:coupledOscillatorsLAndC:diagram}.
Kirchhoff's laws for this circuit give us four equations
\begin{align}
  I_{C_a} + I_{L_a} + I_g &= 0 \nonumber \\
  I_{C_b} + I_{L_b} - I_g &= 0 \nonumber \\
  L_a \dot I_{L_a} + L_g \dot I_{L_b} &= V_a \nonumber \\
  L_b \dot I_{L_b} + L_g \dot I_{L_a} &= V_b
  \, .
\end{align}
Note that the sense of the mutual inductance $L_g$ is defined such that a positive value of $\dot{I}_a$ induces a positive contribution to $V_b$, and vice versa.
Going through the usual process of finding the Lagrangian and then converting to the Hamiltonian, we find the Hamiltonian
\begin{align}\label{eq:ham_coupled_res}
  H_g / \hbar
  &= \omega_a' \abs{a}^2 + \omega_b' \abs{b}^2 \nonumber \\
  &- g_+ (ab + a^* b^*) + g_- (ab^* + a^* b)
\end{align}
where we defined
\begin{align}
  g_c \equiv \frac{1}{2} \frac{1}{C_g' \sqrt{Z_a' Z_b'}} &\qquad
  g_l \equiv \frac{1}{2} \frac{\sqrt{Z_a' Z_b'}}{L_g'} \nonumber \\
  \text{and} \qquad \qquad
  g_+ \equiv g_c + g_l &\qquad g_- \equiv g_c - g_l
\end{align}
where $\omega'$, $Z'$, $C'$, and $L'$ are normalized frequency, impedance, capacitance, and inductance of the coupled resonators.
Details are given in Appendix \ref{appendix:coupling}.
Hamilton's equations of motion can now be expressed in matrix form
\begin{equation*}
  \frac{d}{dt}
  \left( \begin{array}{c} a \\ b \\ a^* \\ b^* \end{array} \right)
  = -i \underbrace{\left( \begin{array}{cccc}
    \omega_a' & g_- & 0 & -g_+ \\
    g_- & \omega_b' & -g_+ & 0 \\
    0 & g_+ & -\omega_a' & -g _- \\
    g_+ & 0 & -g_- & -\omega_b'
  \end{array} \right)}_M
  \left( \begin{array}{c} a \\ b \\ a^* \\ b^* \end{array} \right) \, .
\end{equation*}
There are two \emph{aspects} to the coupling.
The coupling $g_-$ connects $a$ to $b$ and $a^*$ to $b^*$, i.e. it acts within the upper and lower diagonal 2x2 sub-blocks of $M$.
On the other hand, the coupling $g_+$ couples $a$ to $b^*$ and $b$ to $a^*$ as it lies on the anti-diagonal of $M$.
The structure of $M$ and the roles of the two aspects of coupling can be clearly seen by writing $M$ in algebraic form as
\begin{equation}
  M = -i \left[
    \sigma_z \otimes
      \left(
        g_- \sigma_x + \frac{\Delta}{2} \sigma_z + \frac{S}{2} \mathbb{I}
      \right)
    -i g_+ (\sigma_y \otimes \sigma_x)
  \right]
\end{equation}
where $S \equiv \omega_a' + \omega_b'$ and $\Delta \equiv \omega_a' - \omega_b'$.

\leveldown{Rotating wave approximation}

In the rotating frame, defined by $\bar{a}(t) = a(t) \exp(i \omega_a' t)$ and $\bar{b}(t) = b(t) \exp(i \omega_b' t)$, the equation of motion for $\bar{a}(t)$ is
\begin{equation}
  \dot{\bar{a}}(t) = -i
  \left(
    g_- \bar{b} e^{-i \left( \omega_b' - \omega_a' \right)t} - g_+ \bar{b}^* e^{i \left( \omega_a' + \omega_b' \right)t}
  \right)
  \, .
\end{equation}
The time dependence of the first term on the right hand side is much slower than the second term, and just as in the case of the driven resonator, the fast term is dropped under the RWA.
Dropping the fast terms in the equations of motion is equivalent to dropping the $ab$ and $a^* b^*$ terms of the coupling Hamiltonian, i.e. dropping the terms of $M$ proportional to $g_+$ (the antidiagonal).
Therefore, if we engineer the coupling such that $g_+$ is identically zero, then the RWA will be exact.
The condition $g_+ = 0$ means $g_c = -g_l$, i.e. the capacitive and inductive coupling must be of equal magnitude but opposite sign, which we call ``balanced coupling''.
Balanced coupling can also be expressed as the condition $-L'_g / C'_g = Z_a' Z_b'$, i.e. the coupling impedance is equal to minus the geometric mean of the impedances of the two resonators.
Recall that a negative value for $L_g$ means that the mutual inductance must be arranged so that a positive value of $\dot{I}_a$ produces a negative contribution to $V_b$.
If the capacitive coupling is positive, then balanced coupling requires the mutual inductance to be negative.

The antidiagonal of $M$ corresponds to the $\sigma_y \otimes \sigma_x$ term in $M$'s algebraic representation.
In the matrix, we see that dropping these terms decouples the clockwise and counterclockwise rotating modes, so the RWA drops the part of the dynamics coupling the clockwise-rotating modes to the counterclockwise-rotating modes.
Appendix \ref{appendix:eigenvalues} compares the full eigenvalues of $M$ with the eigenvalues of $M$ under the RWA.

\levelstay{Readout}

Balanced coupling may have applications in readout of superconducting qubits.
Superconducting qubits are typically measured via dispersive coupling to a harmonic resonator \cite{Schuster:ac_Stark_dephasing:2005, Wallraff:unit_visibility:2005,Jeffrey:fast_readout:2014, Heinsoo:readout:2018}.
Unfortunately, introducing even modest levels of energy into the resonator during the readout activates resonances in the coupled qubit-resonator system that kick the qubit into noncomputational states high above the transmon potential \cite{Sank:rotating_wave:2016, khezri:2022:within_rotating_wave, Shillito:ionization:2022, Dumas:2024:ionization, Cohen:chaos:2023}.
This effect, called measurement-induced state transitions (MIST), is poisonous to quantum error correction \cite{Miao:DQLR:} as occupation of noncomputational states leads to correlated errors and is difficult to reset.
MIST is a major challenge for quantum computing with superconducting qubits.

Critically, in the usual case where the resonator frequency is larger than the qubit frequency, the resonances responsible for MIST are mediated precisely by the excitation nonpreserving ``non-RWA'' terms in the coupling Hamiltonian, i.e. the terms that are neglected in the RWA \cite{Khezri:dressed_squeezed_states:2016, Sank:rotating_wave:2016}.
Therefore, in a circuit with balanced coupling between the qubit and resonator, MIST may be partially suppressed.
On the other hand, we should not expect perfect suppression because the preceding analysis is for two coupled harmonic resonators while qubits are by definition anharmonic.

To see whether or not MIST can be removed with balanced coupling, we study the matrix elements of the transmon-resonator coupling Hamiltonian.
A perturbative expansion of the coupling is given in Appendix~\ref{appendix:nonlinear_oscillators}.
Using that expansion, we calculate the matrix elements connecting the initial state $\ket{k, n}$ with final states $\ket{k+1, n+1}$ and $\ket{k+3, n-1}$, each having two excitations more than the initial state and therefore requiring non-RWA coupling terms.
These particular final states are of interest because they may serve as intermediate virtual states in one of the two MIST transitions identified in Ref.~\cite{Sank:rotating_wave:2016}.
Results are shown in Fig.~\ref{fig:matrix_elements}.
Interestingly, the matrix element for $\ket{k+1, n+1}$ vanishes when $g_l = -0.9 \, g_c$, close to the balanced coupling condition discussed above for coupled harmonic resonators.
The matrix element for $\ket{k+3, n-1}$ vanishes for $g_l = 3 \, g_c$.
However, there is no ratio $g_l / g_c$ for which both matrix elements vanish simultaneously.
We emphasize that these ratios are calculated using a perturbative treatment.

\quickfig{\columnwidth}{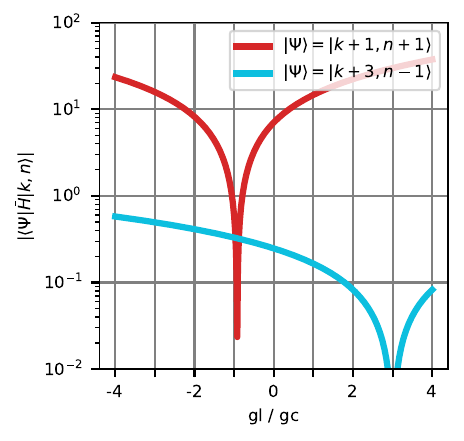}{
Perturbative calculation of two of the matrix elements taking the system from state $\ket{k, n}$ to $\ket{k+1, n+1}$ and $\ket{k+3, n-1}$.
The curves here are for the case $k=3$, $n=20$, $g_c / 2\pi = 180 \, \text{MHz}$, and $\lambda = E_C / 3 \sqrt{8 E_J E_C} = 0.013$.
}{fig:matrix_elements}

We numerically simulate the semiclassical model described in Ref.~\cite{khezri:2022:within_rotating_wave}  but here tailored to account for non-RWA interactions.
The simulation proceeds in two steps.
We first simulate the driven resonator under the influence of a classical drive, where the resonator forms a coherent state.
We then treat the resonator field as a classical drive applied to the qubit~\cite{khezri:2022:within_rotating_wave, Dumas:2024:ionization}.
The qubit-resonator coupling is $g = (\theta/2) \sqrt{\omega_q \omega_r}$ with coupling efficiency $\theta \approx 0.05$.
The resonator frequency is $\omega_r = (2\pi)\,6\,\text{GHz}$ and the resonator linewidth is $\kappa = 1/(15 \, \text{ns})$.
The qubit frequency is set to a value between $4\,\text{GHz}$ and $5\,\text{GHz}$, with transmon charging energy of $E_C = (2\pi)\,0.2\,\text{GHz}$.
We perform the simulation once with balanced coupling and a fixed total coupling efficiency and then again with purely capacitive coupling, but we adjust the latter coupling to achieve the same dispersive shift as in the balanced case at each qubit frequency.
Keeping the same dispersive shift ensures a meaningful comparison of the readout performance at each frequency.
We also repeat the simulation for a set of 20 equally spaced transmon gate offset charges $n_g$ in the range $n_g \in [-0.5, 0]$ and average the results.

\quickwidefig{2\columnwidth}{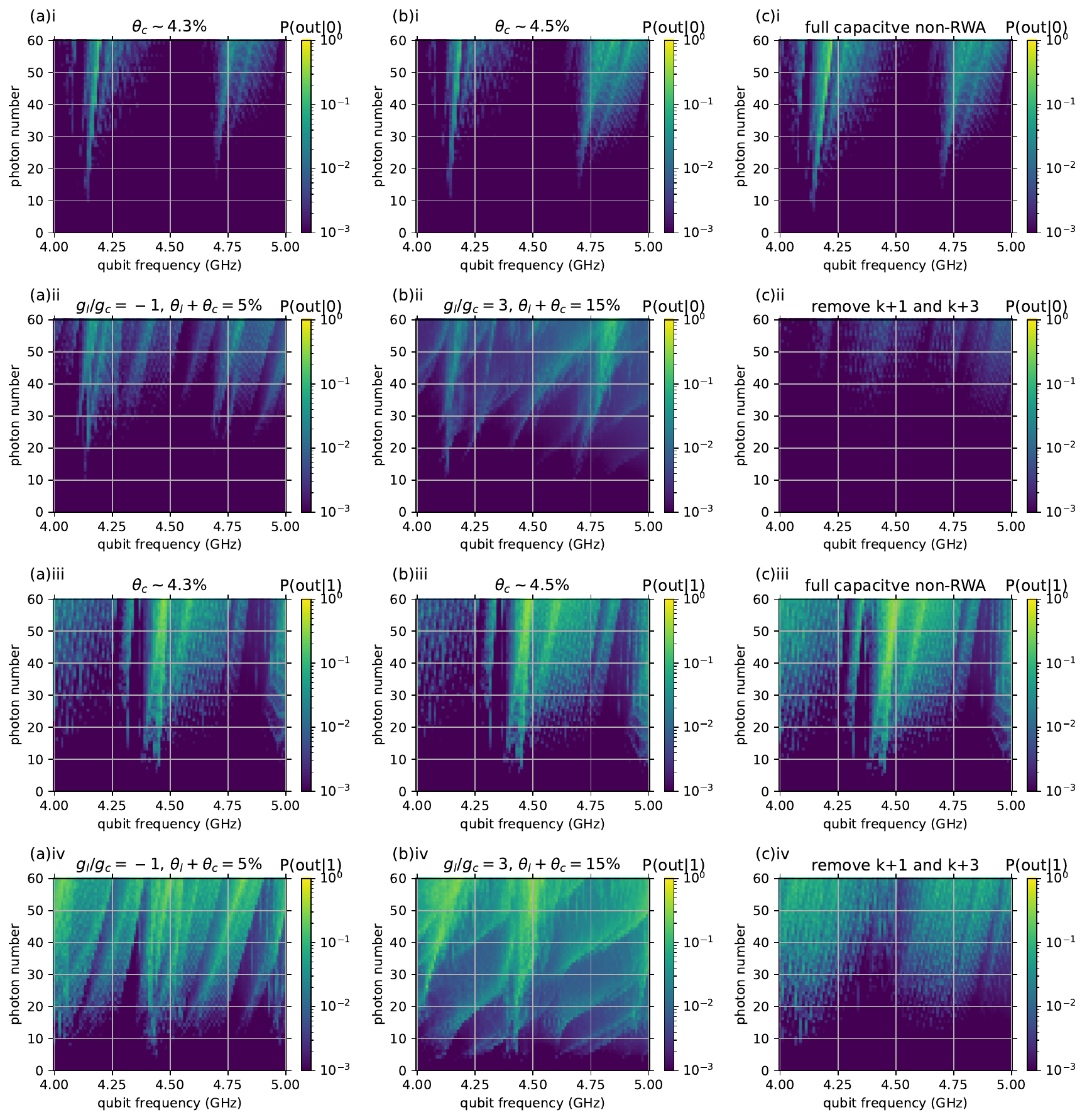}{
Simulations of MIST for the coupled qubit-resonator with and without balanced coupling, for different $g_l/g_c$ ratios.
First two rows show the result for qubit prepared in $\ket{0}$, and the second two rows for qubit prepared in $\ket{1}$.
\textbf{(a)} Simulations for balanced coupling with $g_l / g_c = -1$, with total balanced coupling efficiency of 0.05.
Panels i and iii show the capacitive case, and panel ii and iv show the balanced case.
\textbf{(b)} Simulations for balanced coupling with $g_l / g_c = 3$, with total balanced coupling efficiency of 0.15.
The large coupling efficiency is chosen to roughly yield the same dispersive shift as in the case of panels (a)i-iv.
Panels i and iii show the capacitive case, and panel ii and iv show the balanced case.
\textbf{(c)} Simulations for capacitive only case with capacitive coupling efficiency of 0.05.
Panels i and iii show the case were we retain all the non-RWA terms in the Hamiltonian.
Panels ii and iv show the case were we artificially remove the non-RWA corresponding to transmon charge matrix elements $\bbraket{k+1}{Q}{k}$ and $\bbraket{k+3}{Q}{k}$.
}{fig:balanced}

Results are shown in Fig.~\ref{fig:balanced}.
Panels (a)i-iv show the results where, for the balanced case we take $g_l / g_c = -1$, which would numerically cancel the non-RWA terms stemming from the $\bbraket{k+1}{Q}{k}$ transmon charge matrix elements.
Similarly, panels (b)i-iv show the results where, for the balanced case we take $g_l / g_c = 3$, which would numerically cancel the non-RWA terms stemming from the $\bbraket{k+3}{Q}{k}$ transmon charge matrix elements.
Unfortunately in both of these cases, balanced coupling does not remedy MIST, and in fact makes it worse.
In panels (c)i and iii, we simulate a typical purely capacitive coupling case with all of its naturally occurring non-RWA terms, and in panels (c)ii and iv we artificially remove both of the non-RWA terms that stem from the transmon charge matrix elements mentioned above.
For the initial state $\ket{0}$ we see broad improvement.
For the initial state $\ket{1}$ a large feature near qubit frequency $4.5\,\text{MHz}$ is removed when we remove the matrix elements, but the rest of the features remain more or less the same.
These results suggest that besides the main two terms considered here, other non-RWA terms also play an important role in creating MIST.

\levelstay{Anti-balanced coupling}

We note briefly that in the case $g_l = g_c$ the photon hopping term in the Hamiltonian \eqref{eq:ham_coupled_res} vanishes and only the two-mode squeezing terms remain.
This effect was used to engineer a nonlinear cross-Kerr interaction in Ref.~\cite{Kounalakis:kerr:2018}.
\levelup{Coupled transmission lines} \label{sec:coupled_lines}

\quickfig{\columnwidth}{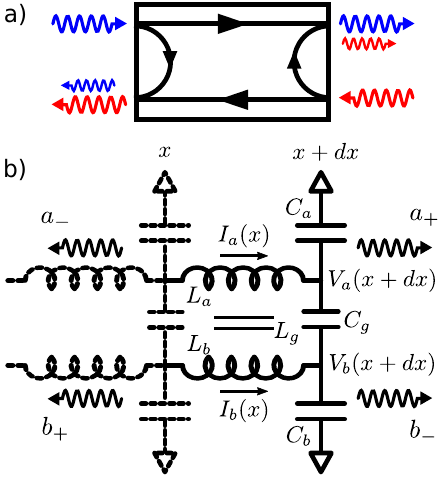}
{a) A directional coupler.
A wave incident on the top left port (blue) is mostly transmitted to the top right port  (large blue), while a small fraction is coupled into the bottom left port (small blue).
Similarly, a wave injected into the bottom right port (red) is mostly transmitted to the bottom left port (large red) while a small fraction is coupled into the top right port (small red).
b) Capacitively and inductively coupled transmission lines.
Each line is modelled as an infinite ladder of lumped capacitors and inductors. Coupling capacitance per length $C_g$ and coupling inductance per length $L_g$ couple the two lines.
}
{fig:directional_coupler}

Our investigation into electromagnetic resonators with simultaneous electric and magnetic coupling was motivated by an attempt to gain intuition into why a directional coupler works.
A directional coupler is a four port microwave device that directs signal flow as shown in Fig.~\ref{fig:directional_coupler}\,a.
A wave injected into the top left port is mostly transmitted through to the top right port, while a small fraction is coupled to the lower left port and none is coupled to the bottom right port.
The key element in a directional coupler is a pair of coupled transmission lines which we model as an infinite ladder of lumped capacitances and inductances as illustrated in Fig.~\ref{fig:directional_coupler}\,b.
The coupled lines act as a directional coupler if the right-moving wave in the top line, represented by $a_+$ and analogous to the blue wave in Fig.~\ref{fig:directional_coupler}, couples only to the left moving wave in the bottom line, represented by $b_+$.

Line $a$ has inductance and capacitance \emph{per length} $L_a$ and $C_a$, and similarly for line $b$.
The lines have characteristic impedances $Z_a$ and $Z_b$.
The lines are coupled by a coupling inductance and capacitance per length $L_g$ and $C_g$.
We define amplitudes $a$ and $b$ by
\begin{align}
  a_\pm \equiv & \frac{V_a}{\sqrt{Z_a}} \pm \sqrt{Z_a} I_a \nonumber \\
  b_\pm \equiv & \frac{V_b}{\sqrt{Z_b}} \mp \sqrt{Z_b} I_b
  \, .
\end{align}
Note the similarity to the definitions of $a$ and $b$ in the case of coupled resonators.
Assuming a sinusoidal time dependence with frequency $\omega$, Kirchhoff's laws for the coupled line circuit can be expressed in matrix form
\begin{equation}
  \frac{d}{dx}
  \begin{pmatrix}
    a_+ \\ b_+ \\ a_- \\ b_-
  \end{pmatrix}
  = i
  \begin{pmatrix}
    \beta_a & -\chi & 0 & \kappa \\
    \chi & -\beta_b & -\kappa & 0 \\
    0 & - \kappa & -\beta_a & \chi \\
    \kappa & 0 & -\chi & \beta_b
  \end{pmatrix}
  \begin{pmatrix}
    a_+ \\ b_+ \\ a_- \\ b_-
  \end{pmatrix}
\end{equation}
where $\beta = \omega / v$ is the wave vector and
\begin{align}
  \kappa =& \frac{\omega}{2}
    \left( \frac{L_g}{\sqrt{Z_a Z_b}} - C_g \sqrt{Z_a Z_b} \right) \nonumber \\
  \chi =& \frac{\omega}{2}
    \left( \frac{L_g}{\sqrt{Z_a Z_b}} + C_g \sqrt{Z_a Z_b} \right)
  \, .
\end{align}
See Appendix \ref{appendix:coupled_lines} for details.
Note the resemblance between the matrix here and the matrix $M$ that arose in the analysis of coupled resonators: each has an approximately block-diagonal form where the anti-diagonal contains only a single parameter, in this case $\kappa$.
Waves $a_+$ and $b_-$ decouple and the lines behave as a directional coupler if $\kappa=0$, which ocurs when
\begin{equation}
  Z_g \equiv \sqrt{\frac{L_g}{C_g}} = \sqrt{Z_a Z_b}
\end{equation}
i.e. when the coupling impedance is equal to the geometric mean of characteristic impedances of the two lines, in direct analogy with the case of coupled resonators.
The travelling wave amplitudes here are analogous to the rotating mode amplitudes in the case of coupled resonators and the same electric-magnetic balance required to make the RWA exact in the case of coupled resonators is required to make the pair of coupled transmission lines act as a directional coupler.

The mathematical similarity of the coupled resonators and coupled transmissions lines leads to an interesting insight about the design of directional couplers.
As a pair of coupled transmission lines acts as a directional coupler when the RWA is exact, and as the RWA is generally a good approximation, a pair of coupled transmission lines will generally make a good approximation to a directional coupler, as long as the coupling is sufficiently weak.

\levelstay{Conclusions} \label{sec:conclusions}

We have shown that the rotating wave approximation can be made exact for both driven and coupled electromagnetic resonators.
In each case, the rotating wave approximation is exact when the strengths of the couplings, associated to the two conjugate degrees of freedom (flux and charge), are equal.
Balancing flux and charge drives suppresses nutation in Rabi oscillations when the Rabi rate is a significant fraction of the Larmor precession frequency. 
While MIST is modified when the qubit-resonator coupling is balanced, there is no clear choice of electric and magnetic couplings that remove the leakage for a transmon.
It may be interesting to investigate whether balanced qubit-resonator coupling can remove MIST in other superconducting qubit species.
\levelstay{Acknowledgements}

We thank Alexander Korotkov and Dvir Kafri for illuminating discussions, and we thank Andre Petukhov for encouragement.

\appendix
\levelstay{Review of the RWA} \label{appendix:overview}

The phrase ``rotating wave'' comes from the case of a spin system driven by fields truly rotating in three dimensional space.
In the case of driven and coupled electromagnetic resonators where geometrically rotating fields do not exist, the notion of ``rotating waves'' appears by mathematical analogy to the case of the spin.

Consider a magnetic moment represented by a unit vector $\vec{\rho} = (\rho_x, \rho_y, \rho_z)$.
If the moment is subjected to a magnetic field $\vec{\Omega} = \gamma \vec{B} = (\Omega_x, \Omega_y, \Omega_z)$ \footnote{$\vec{\Omega} = \gamma \vec{B}$ where $\gamma$ is the gyromagnetic ratio, so $\vec{\Omega}$ has dimensions of frequency.} then it precesses according to 
\begin{align*}
  \frac{d}{dt} \vec{\rho} = \vec{\rho} \times \vec{\Omega}
  \, .
\end{align*}
In the typical case, the magnetic field is dominated by a large static component $\vec{\Omega}_0$ which we take to lie along the z-axis, $\vec{\Omega}_0 = \Omega_0 \hat{z}$.
In the presence of just this static field and taking $\vec{\rho}(0) = \hat{x}$, the solution is
\begin{equation}
  \vec \rho (t) = \cos(\Omega_0 t) \hat{x} - \sin(\Omega_0 t) \hat{y} \, ,
  \label{eq:moment_free_motion}
\end{equation}
i.e. the moment rotates clockwise about the axis of the static magnetic field with frequency proportional to the field strength.

The moment can be visualized as a point on a sphere with coordinates $\theta$ and $\phi$ such that $\rho_x = \sin\theta \cos\phi$, $\rho_y = \sin\theta \sin\phi$, and $\rho_z = \cos\phi$.
Suppose that we wish to control the angle $\theta$ by application of an additional drive field $\vec{\Omega}_d(t)$.
In the absence of $\vec{\Omega}_0$ we could rotate $\vec\rho$ about the x-axis with a static field oriented along $\hat{x}$, but when $\Omega_0 \neq 0$ precession about the z-axis causes the relative azimuth angle between $\vec\rho$ and $\hat{x}$ to oscillate in time.
To get a pure rotation of $\theta$ when $\Omega_0 \neq 0$, the axis of the drive field needs to rotate along with the precession induced by $\vec{\Omega}_0$, i.e.
\begin{equation*}
  \vec{\Omega}_d(t)
  \propto \cos(\phi_d + \Omega_0 t) \hat{x} - \sin(\phi_d + \Omega_0 t) \hat{y}
  \, .
\end{equation*}
In the coordinate frame rotating with the precession induced by $\vec{\Omega}_0$, the time dependence of $\vec{\Omega}_d$ vanishes and we're left with (the subscript $r$ indicates the rotating coordinate frame)
\begin{equation}
  \vec{\Omega}_{d, r} \propto \cos(\phi_d) \hat{x} + \sin(\phi_d) \hat{y}
  \label{eq:moment_rotating_drive}
\end{equation}
corresponding to a rotation about an axis in the xy plane with azimuth angle $\phi_d$.

Frequently in real life situations, we don't have access to true rotating fields.
Instead, we may have a linearly polarized field
\begin{equation*}
  \vec{\Omega}_{d,\text{linear}} \propto \cos(\Omega_0 t + \phi_d) \hat{x}
  \, .
\end{equation*}
The linear field can be expressed as a \emph{pair} of rotating fields, one rotating along with the rotating frame and one rotating against the rotating frame:
\begin{align*}
  \vec{\Omega}_{d, \text{linear}}
  &\propto \frac{1}{2} \underbrace{
    \left( \cos(\Omega_0 t + \phi_d) \hat{x} - \sin(\Omega_0 t + \phi_d) \hat{y} \right)
  }_\text{with frame} \\
  &+ \frac{1}{2} \underbrace{
    \left( \cos(\Omega_0 t + \phi_d) \hat{x} + \sin(\Omega_0 t + \phi_d) \hat{y} \right)
  }_\text{against frame}
\end{align*}
In the rotating frame, these fields become
\begin{align*}
  \vec{\Omega}_{d, \text{linear}, r}
  \propto& \frac{1}{2} \underbrace{
    \left( \cos(\phi_d) \hat{x} + \sin(\phi_d) \hat{y} \right)
    }_\text{static} \\
    +& \frac{1}{2} \underbrace{
      \left( \cos(2 \Omega_0 t + \phi_d) \hat{x} + \sin(2 \Omega_0 t + \phi_d) \hat{y} \right)
    }_\text{counter-rotating}
  \, .
\end{align*}
The static part is exactly half of what we had with a true rotating field, but now there's an extra counter-rotating term rotating at frequency $2 \Omega$ in the direction opposite the free precession induced by $\vec{\Omega_0}$.
This inconvenient counter-rotating term is precisely what is neglected in the RWA on the grounds that, because it rotates at such a high frequency, its influence on the dynamics integrates to nearly zero.
When the RWA is used to drop the counter-rotating term from $\vec{\Omega}_{d, \text{linear}, r}$, then $\vec{\Omega}_{d, \text{linear}, r} \propto \vec{\Omega}_{d,r}$.
In the context of a magnetic moment suspended in space, the meaning of the term ``rotating wave approximation'' is clear as the RWA essentially replaces a linearly polarized drive field with a rotating one.
With that in mind, it may seem that the use of the RWA in the context of electrical circuits, where there is no real three dimentional space in which to construct rotating drive fields, would be inevitable.
However, as shown in the main text, drives completely analogous to $\vec{\Omega}_d(t)$ can be designed in electrical circuits by proper balance of electric and magnetic aspects of the drive.

\section{Driving} \label{appendix:driving}

\leveldown{Kirchhoff's laws}

Kirchhoff's laws for the driving circuit shown in Figure \ref{fig:driven_resonator} are
\begin{align}
  C_d \left( \ddot{V}_d - \ddot{V} \right) &= \dot{I}_C + \dot{I}_L \nonumber \\
  L \dot{I}_L + M_d \dot{I}_d &= V \, .
\end{align}
These two equations can be combined into a single differential equation
\begin{equation*}
  \ddot{V} \left(1 + \frac{C_d}{C} \right) + \omega_0^2 V = \frac{C_d}{C} \ddot{V}_d + \omega_0^2 M_d \dot{I}_d \, .
\end{equation*}
Defining $\Phi = \int V dt$, $C' = C + C_d$, and $\omega_0' = \omega_0 / \sqrt{1 + C_d / C}$, the equation can be rewritten as
\begin{equation}
  \ddot{\Phi} + \omega_0'^2 \Phi = \frac{C_d}{C'} \dot{V}_d(t) + \omega_0'^2 M_d I_d(t)
  \, .
\end{equation}

\levelstay{Hamiltonian form}

This equation is recovered by the Lagrangian
\begin{align}
  \mathcal{L} 
  &= \frac{1}{2} C \dot{\Phi}^2 - \frac{1}{2L} \Phi^2 \nonumber \\
  &+ \frac{1}{2} C_d \left( \dot{\Phi} - V_d(t) \right)^2 + \frac{M_d}{L} I_d(t) \Phi
  \, .
\end{align}
The momentum conjugate to $\Phi$ is
\begin{equation}
  \frac{\partial \mathcal{L}}{\partial \dot{\Phi}} = C' \dot{\Phi} - C_d V_d(t) \equiv Q
\end{equation}
and the Hamiltonian is
\begin{align}
  H
  &= Q \dot{\Phi} - \mathcal{L} \\
  &= \frac{Q^2}{2 C'} + \frac{\Phi^2}{2L} \nonumber \\
  &+ \frac{C_d}{C'} V_d(t) Q - \frac{M_d}{L} I_d(t) \Phi
\end{align}
which is where we begin the analysis in the main text.

\levelstay{Two-level system}

The analysis in the main text is entirely classical with $\hbar$ an arbitrary constant with dimensions of action.
However, an important case is when the driven resonator is both anharmonic and quantum mechanical, such as with superconducting qubits.
If the drive frequency is set on or near resonance for only a single quantum transition $\ket{n} \rightarrow \ket{m}$, then the projection of $H_d$ (see Eq.~(\ref{eq:drive_hamiltonian_lab_frame})) onto the two-level subspace spanned by those two states is approximately\footnote{This approximation leaves out terms proportional to $\sigma_z$ which can usually be ignored under assumptions similar to the RWA.}
\begin{align}
  H_d
  =&
      \frac{C_d}{C_\Sigma} V_d(t) \abs{Q_{n,m}} \sigma_y
    - \frac{L_d}{L} I_d(t) \abs{\Phi_{n,m}} \sigma_x \nonumber \\
  =& - z(t) \sigma_- - z(t)^* \sigma_+
  \, .
\end{align}
where $Q_{n,m} = \bbraket{n}{Q}{m}$, $\Phi_{n,m} = \bbraket{n}{\Phi}{m}$, and $z(t)$ has the same meaning as before except that now
\begin{align*}
  G_x =& \, (L_d / L) \abs{\Phi_{n,m}} I_d \\
  G_y =& \, (C_d / C_\Sigma) \abs{Q_{n,m}} V_d
  \, .
\end{align*}
With the drive $z(t) = G_d \exp(i (\omega_d t + \phi_d)$ and in the frame rotating at frequency $\omega_d$,
\begin{align}
  H_{d,r}
  =& - G_d \left( e^{-i \phi_d} \sigma_+ + e^{i \phi_d} \sigma_- \right) \nonumber \\
  =& - G_d \begin{pmatrix}
    0 & e^{i \phi_d} \\ e^{-i \phi_d} & 0
  \end{pmatrix} \nonumber \\
  =& - G_d \left( \cos(\phi_d) \sigma_x - \sin(\phi_d) \sigma_y \right)
\end{align}
which is completely analogous to Eq.~(\ref{eq:moment_rotating_drive})
This result has a simple and well known interpretation:
a two-level quantum system acts like a magnetic moment (a point on the Bloch sphere) and in the rotating frame, a drive resonant with that two-level system's transition acts like two perpedicular components of a ficticious three dimensional drive field in the xy-plane.

\section{Coupling} \label{appendix:coupling}

\leveldown{Kirchhoff's laws}

In order to get useful equations of motion from Kirchhoff's laws, we need to work with all voltages or all currents.
We'll use voltages.
The capacitor currents are easily related to voltage via the constitutive equation for a capacitor: $C_i \dot V_i = I_i$.
Note that for the coupling capacitor, the constitutive relation gives $I_g = C_g (\dot V_a - \dot V_b)$.
Relating the inductor currents to voltages requires more work.
The bottom two of the equations from Kirchhoff's laws can be written in matrix form as
\begin{equation*}
  \begin{pmatrix} V_a \\ V_b \end{pmatrix}
  = \underbrace{ \begin{pmatrix}
    L_a & L_g \\
    L_g & L_b
  \end{pmatrix}}_{T_L}
  \begin{pmatrix} \dot I_a \\ \dot I_b \end{pmatrix}
  \, .
\end{equation*}
The inverse of $T_L$ is
\begin{align*}
  T_L^{-1}
  &= \frac{1}{L_a L_b - L_g^2} \begin{pmatrix}
    L_b & -L_g \\ -L_g & L_a
  \end{pmatrix} \\
  & \equiv \begin{pmatrix}
    1 / L_a' & -1 / L_g' \\ -1 / L_g' & 1 / L_b'
  \end{pmatrix}
\end{align*}
so that
\begin{align*}
  \dot I_{L_a} &= \frac{V_a}{L_a'} - \frac{V_b}{L_g'} \\
  \dot I_{L_b} &= \frac{V_b}{L_b'} - \frac{V_a}{L_g'} \, .
\end{align*}
Note that
\begin{align*}
  L_a' & \equiv L_a - \frac{L_g^2}{L_b} \\
  \text{and} \quad
  L_b' & \equiv L_b - \frac{L_g^2}{L_a}
\end{align*}
are the inductances to ground for each resonator, including the inductance through the mutual.
Now we can rewrite all of the currents in the first two of Kirchhoff's laws entirely in terms of $V_a$ and $V_b$:
\begin{align*}
  C_a \ddot V_a + \frac{V_a}{L_a'} - \frac{V_b}{L_g'} + C_g (\ddot V_a - \ddot V_b) &= 0 \\
  C_b \ddot V_b + \frac{V_b}{L_b'} - \frac{V_a}{L_g'} + C_g (\ddot V_b - \ddot V_a) &= 0 \, .
\end{align*}
Traditional analysis of circuits in the physics literature uses flux and charge instead of current and voltage, so defining $\Phi = \int V \, dt$, we can write our equations of motion as
\begin{align}
  \ddot \Phi_a (C_a + C_g) - C_g \ddot \Phi_b + \frac{\Phi_a}{L_a'} - \frac{\Phi_b}{L_g'} &= 0 \nonumber \\
  \ddot \Phi_b (C_b + C_g) - C_g \ddot \Phi_a + \frac{\Phi_b}{L_b'} - \frac{\Phi_a}{L_g'} &= 0
  \label{eq:coupled_equations_of_motion}
  \, .
\end{align}

\levelstay{Hamiltonian form}

By inspection and a bit of fiddling around, one can check that equations (\ref{eq:coupled_equations_of_motion}) are generated by the Lagrangian
\begin{align}
  \mathcal{L}
  =& \underbrace{\frac{C_g}{2} \left(\dot \Phi_a - \dot \Phi_b \right)^2
   + \frac{C_a}{2} \dot \Phi_a^2 + \frac{C_b}{2} \dot \Phi_b^2}_\text{kinetic} \nonumber \\
  & \underbrace{- \frac{\Phi_a^2}{2 L_a'} - \frac{\Phi_b^2}{2 L_b'} + \frac{\Phi_a \Phi_b}{L_g'}}_\text{potential} \, .
\end{align}
The canonical momenta conjugate to $\Phi_a$ and $\Phi_b$ are
\begin{align*}
  Q_a & \equiv \frac{\partial \mathcal{L}}{\partial \dot \Phi_a} = (C_a + C_g) \dot \Phi_a - C_g \dot \Phi_b \\
  Q_b & \equiv \frac{\partial \mathcal{L}}{\partial \dot \Phi_b} = (C_b + C_g) \dot \Phi_b - C_g \dot \Phi_a
  \, .
\end{align*}
or in matrix form
\begin{equation*}
  \begin{pmatrix} Q_a \\ Q_b \end{pmatrix}
  =
  \underbrace{
    \begin{pmatrix}
       (C_a + C_g) & -C_g \\
       -C_g & (C_b + C_g)
    \end{pmatrix}
  }_{T_C}
  \begin{pmatrix} \dot \Phi_a \\ \dot \Phi_b \end{pmatrix}
  \, .
\end{equation*}
The matrix $T_C$ is just the capacitance matrix of the circuit.
Its inverse is
\begin{align*}
  T_C^{-1} =& \frac{1}{C_a C_b + C_g (C_a + C_b)}
  \begin{pmatrix}
    (C_b + C_g) & C_g \\
    C_g & (C_a + C_g)
  \end{pmatrix} \\
  \equiv& \begin{pmatrix}
    1 / C_a' & 1 / C_g' \\
    1 / C_g' & 1 / C_b'
  \end{pmatrix}
\end{align*}
Note that $C_a'$ and $C_b'$ are simply the total capacitances to ground for each resonator!

The Hamiltonian function $H$ for the system is defined formally by the equation
\begin{equation*}
  H \equiv \left( \sum_{i = \{a, b\}} Q_i \dot \Phi_i \right) - \mathcal{L}
\end{equation*}
where $\Phi_i$ are the coordinates and $Q_i$ are the conjugate momenta, but where we have to replace the $\dot \Phi$'s in both the $Q \dot \Phi$ terms and in $\mathcal{L}$ with $\Phi$'s and $Q$'s.
To do this, first note that the kinetic term in the Lagrangian can be expressed as\footnote{Because matrix transposition and inversion commute, and because the matrix $T_C$ is symmetric, we can bring $T_C^{-1}$ from the bra onto the ket for free.}
\begin{align*}
  \mathcal{L}_\text{kinetic}
  =& \frac{1}{2} \bbraket{\dot \Phi}{T_C}{\dot \Phi} \\
  =& \frac{1}{2} \bbraket{T_C^{-1} Q}{T_C}{T_C^{-1} Q} \\
  =& \frac{1}{2} \bbraket{Q}{T_C^{-1}}{Q}
  \, .
\end{align*}
Note also that $\sum_i \dot \Phi_i Q_i = \bbraket{Q}{T_C^{-1}}{Q}$, so we can write the Hamiltonian as
\begin{align}
  H
  =& \left( \sum_{i = \{a, b\}} Q_i \dot \Phi_i \right) - \mathcal{L} \nonumber \\
  =& \bbraket{Q}{T_C^{-1}}{Q} - \left( \mathcal{L}_\text{kinetic} + \mathcal{L}_\text{potential} \right) \nonumber \\
  =& \bbraket{Q}{T_C^{-1}}{Q} - \left( \frac{1}{2} \bbraket{Q}{T_C^{-1}}{Q} + \mathcal{L}_\text{potential} \right) \nonumber \\
  =& \frac{Q_a^2}{2 C_a'} + \frac{Q_b^2}{2 C_b'} + \frac{\Phi_a^2}{2 L_a'} + \frac{\Phi_b^2}{2 L_b'} \nonumber \\
  &+ \underbrace{\frac{Q_a Q_b}{C_g'} - \frac{\Phi_a \Phi_b}{L_g'}}_{\text{coupling Hamiltonian }H_g}
  \, .
\end{align}

\levelstay{Dimensionless variables}

We will now simplify this Hamiltonian so that we can easily find its normal modes.
First, we define
\begin{align*}
  X_i &\equiv \frac{1}{\sqrt{2 \hbar}} \frac{1}{\sqrt{Z_i'}} \Phi_i \\
  Y_i &\equiv \frac{1}{\sqrt{2 \hbar}} \sqrt{Z_i'} Q_i
\end{align*}
where $Z_i' \equiv \sqrt{L_i' / C_i'}$, and we've added the constant $\hbar$ with dimensions of action to make $X$ and $Y$ dimensionless.
This entire analysis has been classical, and in the classical case $\hbar$ can be thought of as \emph{anything} with dimensions of action.
Of course, in the quantum case, we should simply think of $\Phi$, $Q$, $X$, and $Y$ as operators and $\hbar$ as Planck's constant.

In the new coordinates, the Hamiltonian is
\begin{align}
  H / \hbar
  =& \omega_a' \left(X_a^2 + Y_a^2 \right)
  +  \omega_b' \left(X_b^2 + Y_b^2 \right) \nonumber \\
  &+ 2 \frac{1}{C_g' \sqrt{Z_a' Z_b'}} Y_a Y_b
   - 2 \frac{\sqrt{Z_a Z_b}}{L_g'} X_a X_b
\end{align}
where $\omega_i' \equiv \sqrt{L_i' / C_i'}$ are called \textbf{partial frequencies} and play an important role in the analysis of the system, particularly when making approximations.

\levelstay{Rotating modes}

Finally we define
\begin{align}
  a &\equiv X_a + i Y_a \nonumber \\
  b &\equiv X_b + i Y_b
\end{align}
to arrive at
\begin{align}
  H / \hbar
  &= \omega_a' a^* a + \omega_b' b^* b \nonumber \\
  & - \frac{1}{2} \frac{1}{C_g' \sqrt{Z_a' Z_b'}} (ab + a^* b^* - a b^* - a^* b) \nonumber \\
  &- \frac{1}{2} \frac{\sqrt{Z_a' Z_b'}}{L_g'} (ab + a^* b^* + a^* b + a b^*)
  \, .
\end{align}
The stars indictate Hermitian conjugation, which in the classical case reduces to complex conjugation.
The coupling term can be reorganized in a very useful form:
\begin{align}
  H_g / \hbar =
  &- \left( a b + a^* b^* \right)
    \frac{1}{2} \left(
      \frac{1}{C_g' \sqrt{Z_a' Z_b'}} + \frac{\sqrt{Z_a' Z_b'}}{L_g'}
    \right) \nonumber \\
  &+ \left( a b^* + a^* b \right)
    \frac{1}{2} \left(
      \frac{1}{C_g' \sqrt{Z_a' Z_b'}} - \frac{\sqrt{Z_a' Z_b'}}{L_g'}
    \right) \nonumber \\
  &= -g_+ (ab + a^* b^*) + g_- (ab^* + a^* b)
\end{align}
where we defined
\begin{align}
  g_c \equiv \frac{1}{2} \frac{1}{C_g' \sqrt{Z_a' Z_b'}} &\qquad
  g_l \equiv \frac{1}{2} \frac{\sqrt{Z_a' Z_b'}}{L_g'} \nonumber \\
  \text{and} \qquad \qquad
  g_+ \equiv g_c + g_l &\qquad g_- \equiv g_c - g_l
  \, ,
\end{align}
which is the starting point in the main text.

\levelstay{Eigenvalues} \label{appendix:eigenvalues}

\leveldown{Eigenvalues in the rotating wave approximation}

The rotating wave approximation provides a convenient approximation for the normal mode frequencies of the coupled resonator system, given by the eigenvalues of $M$.
In the rotating wave aproximation, the algebraic representation of $M$ reduces to
\begin{equation}
  M_\text{RWA} = -i \sigma_z \otimes \left(
    g_- \sigma_x + \frac{\Delta}{2} \sigma_z + \frac{S}{2} \mathbb{I}
  \right)
  \, .
  \label{eq:matrix_algebra_rwa}
\end{equation}
Equation (\ref{eq:matrix_algebra_rwa}) makes finding the eigenvalues particularly simple because the eigenvalues of a tensor product are the products of the eigenvalues of the individual factors.
The eigenvalues of the quantity in parentheses, and therefore the normal mode frequencies, are
\begin{equation}
  \pm \omega_\pm
  = \pm \left(
    \frac{\omega_a' + \omega_b'}{2} \pm \sqrt{g_-^2 + (\Delta / 2)^2 }
  \right)
  \, .
\end{equation}
These eigenvalues are shown in Figure~\ref{Figure:eigenvalues}\,a as a function of both $g_-$ and $\omega_b'$ where we see the famous avoided level crossing.
The symmetry of Figure~\ref{Figure:eigenvalues}\,a exists because we plot the normal frequencies against the frequency $\omega_b'$ rather than the uncoupled frequency $\omega_b$.

\levelstay{Accuracy of the rotating wave approximation}

How accurate are the RWA eigenvalues plotted in Figure~\ref{Figure:eigenvalues}\,a?
In Figure~\ref{Figure:eigenvalues}\,b We compare the full coupled resonator eigenvalues against the RWA, working in the case that either $g_c$ or $g_l$ is zero so that $g_+ = g_- \equiv g$.
For $g=0.01$ and $g=0.2$ the curves are visually indistinguishable.
For $g=1$ the curves just begin to separate, suggesting that the RWA begins to break down when the coupling is on the order of 10\% of the frequencies of the coupled resonators.
Figure~\ref{Figure:eigenvalues}\,c shows the eigenvalues taken at the avoided level crossing ($\omega_b' = \omega_a'$) and Figure~\ref{Figure:eigenvalues}\,d shows the relative error of the RWA at the avoided level crossing.

\quickwidefig{2\columnwidth}{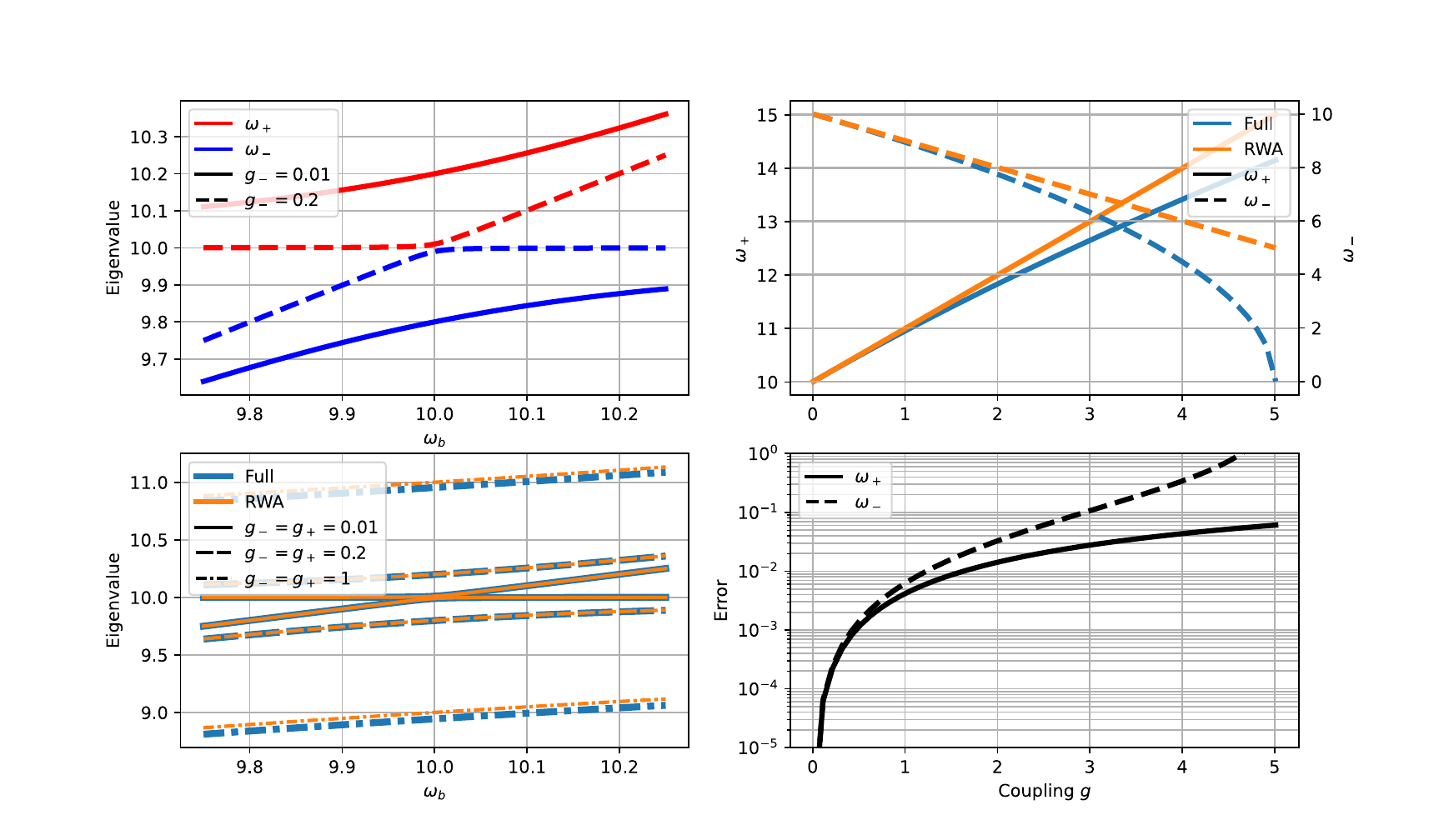}{Eigenvalues of the coupled resonator system. a) Eigenvalues $\omega_\pm$ as a function of $\omega_b'$ for $\omega_a' = 10$. b) Full eigenvalues (i.e. not taking the RWA) compared with the RWA. Here $g_-$ and $g_+$ are taken to be equal as appropriate for resonators coupled through only one of their conjugate variables, i.e. either flux or charge. c) Full and RWA eigenvalues $\omega_\pm$ at the avoided level crossing defined by $\omega_b'=\omega_a'$. d) Relative error of the RWA.}{Figure:eigenvalues}

\levelup{Coupled transmission lines} \label{appendix:coupled_lines}

Kirchhoff's laws for the circuit shown in Fig.~\ref{fig:directional_coupler}\,b are
\begin{align*}
  V_a(x) - V_a(x + dx) =& L_a \dot I_a + L_g \dot I_b \\
  I_a(x) - I_a(x + dx) =& \dot V_a C_a + \left( \dot V_a - \dot V_b \right) C_g \\
  V_b(x) - V_b(x + dx) =& L_b \dot I_b + L_g \dot I_a \\
  I_b(x) - I_b(x + dx) =& \dot V_b C_b + \left( \dot V_b - \dot V_a \right) C_g
  \, .
\end{align*}
In the limit $dx \rightarrow 0$, these equations become differential equations
\begin{align}
  - \frac{\partial V_a}{\partial x} &= \frac{\partial}{\partial t}
    \left( L_a I_a + L_g I_b \right) \nonumber \\
  - \frac{\partial I_a}{\partial x} &= \frac{\partial}{\partial t}
    \left( (C_a + C_g)V_a - C_g V_b \right) \nonumber \\
  - \frac{\partial V_b}{\partial x} &= \frac{\partial}{\partial t}
    \left( L_b I_b + L_g I_a \right) \nonumber \\
  - \frac{\partial I_b}{\partial x} &= \frac{\partial}{\partial t}
    \left( (C_b + C_g)V_b - C_g V_a \right)
  \, .
\end{align}
Assuming sinusoidal time dependence at frequency $\omega$, and therefore converting time derivatives to $-i \omega$, these equations can be written in matrix form as~\footnote{The choice of sign in $-i \omega$ makes positive values of the wave vector correspond to right-moving waves.}
\begin{equation}
  \frac{d}{dx} \left(
    \begin{array}{c} V_a \\ I_a \\ V_b \\ I_b \end{array}
  \right)
  =
  i \omega \left( \begin{array}{cccc}
    0 & L_a & 0 & L_g \\
    C_a & 0 & -C_g & 0 \\
    0 & L_g & 0 & L_b \\
    -C_g & 0 & C_b & 0
  \end{array} \right)
  \left(
    \begin{array}{c} V_a \\ I_a \\ V_b \\ I_b \end{array}
  \right) \, .
\end{equation}
If we convert to the ingoing and outcoming wave amplitudes
\begin{align}
  a_\pm \equiv& \frac{V_a}{\sqrt{Z_a}} \pm \sqrt{Z_a} I_a \nonumber \\
  b_\pm \equiv& \frac{V_b}{\sqrt{Z_b}} \mp \sqrt{Z_b} I_b \, ,
\end{align}
the equations of motion take the form
\begin{equation}
  \frac{d}{dx} \left(
    \begin{array}{c}
      a_+ \\ b_+ \\ a_- \\ b_-
    \end{array}
  \right)
  =
  i \left(
    \begin{array}{cccc}
      \beta_a & -\chi & 0 & \kappa \\
      \chi & - \beta_b & -\kappa & 0 \\
      0 & -\kappa & -\beta_a & \chi \\
      \kappa & 0 & - \chi & \beta_b
    \end{array}
  \right)
  \left(
    \begin{array}{c}
      a_+ \\ b_+ \\ a_- \\ b_-
    \end{array}
  \right)
\end{equation}
where $\beta \equiv \omega / v$ is the wave vector and
\begin{align}
  \kappa &\equiv \frac{\omega}{2} \left( \frac{L_g}{\sqrt{Z_a Z_b}} - C_g \sqrt{Z_a Z_b} \right) \nonumber \\
  \chi &\equiv \frac{\omega}{2} \left( \frac{L_g}{\sqrt{Z_a Z_b}} + C_g \sqrt{Z_a Z_b} \right) \, .
\end{align}

\levelstay{Balanced coupling with nonlinear oscillators} \label{appendix:nonlinear_oscillators}

In the main text we showed that the non-RWA terms $ab$ and $a^\dagger b^\dagger$ can be exactly canceled from the Hamiltonian of two coupled linear oscillators by balancing the electric and magnetic couplings.
In this section we analyze the case where one of the oscillators is a transmon qubit (a weakly nonlinear oscillator), while the other resonator is the qubit readout resonator, assumed to be a linear oscillator.
In contrast to the linear case, we will show that, in the nonlinear case, new non-RWA terms appear that cannot be simultaneously canceled under the balanced coupling condition.
In this section, we use a quantum description throughout.  

We use the following approximate Hamiltonian for the transmon qubit~\cite{MostafaThesis} 
\begin{align}
\label{SM-H_nonlinear_osc}
  H_{\rm q}
  = \hbar \omega_p\left[ b^\dagger b +\frac{1}{2}- \lambda
  \left(\frac{b+b^\dagger}{\sqrt{2}}\right)^4\right],
\end{align}
where $b$ and $b^\dagger$ are the (bare) qubit annihilation and creation operators, respectively.
$\omega_p=\sqrt{8E_cE_J}/\hbar$ is the plasma frequency, where $E_c$ and $E_J$ are respectively the charging and Josephson energies. $\lambda=E_c/(3\hbar\omega_p)$ is a dimensionless parameter characterizing the strength of the qubit nonlinearity ($\lambda>0$).
Typically for a transmon qubit, $\lambda$ is of order of $10^{-2}$.
Here we define the dimensionless charge ($\hat Q_{\rm q}$) and flux ($\hat \Phi_{\rm q}$) operators of the qubit as 
\begin{align}
  Q_{\rm q} = \frac{b-b^\dagger}{i},\;\;   \Phi_{\rm q}=b+b^\dagger.
\end{align}
These dimensionless operators are normalized differently than $X$ and $Y$ in the main text; in particular here $\left[ \Phi_{\rm q}, Q_{\rm q} \right] = 2i$.
The qubit readout resonator is assumed to be linear with Hamiltonian equal to $H_{\rm r} = \hbar \omega_{\rm r} a^\dagger a$, where $a$ and $a^\dagger$ are the photon annihilation and creation operators.
The dimensionless charge and flux operators for the readout resonator are 
\begin{align}
    Q_{\rm r} = \frac{a-a^\dagger}{i},\;\; \Phi_{\rm r}=a+a^\dagger.
\end{align}
The coupling between the two oscillators is similar to the electric-magnetic coupling discussed in the main text, 
\begin{align}
\label{SM-coupling}
  H_{\rm coupling}
  &= \hbar g_c Q_{\rm q} Q_{\rm r} - \hbar g_l \Phi_{\rm q} \Phi_{\rm r}\nonumber \\
  &= -\hbar g_c (b - b^\dagger)(a - a^\dagger) - \hbar g_l (b + b^\dagger)(a+a^\dagger)
  \, .
\end{align}
The balanced coupling condition reads as $g_c = -g_l$. 

The analysis of the non-RWA terms is carried out in the eigenbasis of each oscillator.
We denote the eigenbasis of the transmon qubit by $|k\rangle$, where $k=0,1,2,\ldots$.
We can use perturbation theory to obtain the qubit eigenstates in terms of the bare qubit states $|k\rangle_{\rm bare}$~\cite{MostafaThesis} and determine the unitary operator $U$ that relates them, 
\begin{align}
  \ket{k} = U \ket{k}_{\rm bare} \, .
\end{align}
To first order in $\lambda$, $U$ is given by ($\openone$ is the identity operator)
\begin{align}
  \label{SM-U}
  U =
    \openone
    -\frac{\lambda}{4}\left[\frac{b^4}{4}
    - \frac{{b^\dagger}^4}{4} + b^2 (2bb^\dagger -3)
    - (2bb^\dagger - 3){b^\dagger}^2 \right]
    \, .
\end{align}
Using Eq.~\eqref{SM-U}, the qubit eigenstates $\ket{k}$, to first order in $\lambda$, are given by 
\begin{align}
  \ket{k}
  =& \ket{k}_{\rm bare} -\frac{\lambda}{4}\bigg[\frac{\sqrt{k(k-1)(k-2)(k-3)}}{4}\, \ket{k-4}_{\rm bare} \nonumber \\
   & -\frac{\sqrt{(k+1)(k+2)(k+3)(k+4)}}{4} \, \ket{k+4}_{\rm bare} \nonumber \\
   & + (2k-1)\sqrt{k(k-1)} \, \ket{k-2}_{\rm bare} \nonumber \\
   & - (2k+3)\sqrt{(k+1)(k+2)}\, \ket{k+2}_{\rm bare} \bigg],
\end{align}
which coincides with Eq.~(A.31) of Ref.~\cite{MostafaThesis}. 

In the qubit eigenstate basis, the annihilation operator $b$ transforms into the operator $\bar b$, which, to first order in $\lambda$, is given by
\begin{align}
  \label{SM-b_dressed}
  \bar b
  \equiv U^\dagger b U
  = b +\frac{\lambda}{4}{b^\dagger}^3 -\frac{\lambda}{2}b^3 + \frac{3\lambda}{2} (b^\dagger b)b^\dagger.  
\end{align}
Equation~\eqref{SM-b_dressed} is useful to compute the matrix elements $b_{kk'}=\bbraket{k}{b}{k'}$ as the latter can be rewritten
\begin{equation}
  b_{kk'}
  =\,_{\rm bare}\bbraket{k}{U^\dagger b U}{k'}_{\rm bare}
  =\,_{\rm bare}\bbraket{k}{\bar b}{k'}_{\rm bare}
\end{equation}
Similarly, in the qubit eigenstate basis, the charge and flux operators are
\begin{align}
  \bar{\Phi}_{\rm q}
  &\equiv U^\dagger \Phi_{\rm q} U = \bar b + {\bar b}^\dagger \nonumber \\
  &= b + b^\dagger - \frac{\lambda}{4}\left(b^3 + {b^\dagger}^3\right) + \frac{3\lambda}{2}\left(b\cdot b^\dagger b + b^\dagger b\cdot b^\dagger\right), \label{SM-flux_operator_dressed} \\
  i\bar{Q}_{\rm q}
    &\equiv i \, U^\dagger Q_{\rm q} U = \bar b - {\bar b}^\dagger \nonumber \\
    &= \left[b - b^\dagger - \frac{3\lambda}{4}\left(b^3 - {b^\dagger}^3\right) - \frac{3\lambda}{2}\left(b\cdot b^\dagger b - b^\dagger b\cdot b^\dagger\right) \right], \label{SM-charge_operator_dressed}
\end{align} 
and the coupling Hamiltonian becomes 
\begin{align}
  \bar H_{\rm coupling}
  =& U^\dagger H_{\rm coupling} U \nonumber \\
  =& \hbar \left( -g_l \bar{\Phi}_{\rm q} - i g_c \bar{Q}_{\rm q} \right) a + \hbar \left( -g_l \bar{\Phi}_{\rm q} + i g_c \bar{Q}_{\rm q} \right)a^\dagger
  \label{SM-coupling_dressed}
  \, .
\end{align}

Let us assume the balanced coupling condition holds, i.e. $g_c = -g_l = g / 2$.
After inserting Eqs.~\eqref{SM-flux_operator_dressed}--\eqref{SM-charge_operator_dressed} into Eq.~\eqref{SM-coupling_dressed}, we find that the coupling Hamiltonian still includes non-RWA terms and can be written as
\begin{align}
  H_{\rm coupling}^{\rm (balanced)} = H_{\rm RWA} + H_{\rm non-RWA}, 
\end{align}
where $$H_{\rm RWA} = g(ab^\dagger + a^\dagger b)$$ is the desired RWA coupling between the eigenstates $\ket{k,n}$ and $\ket{k \mp 1,n \pm 1}$, belonging to the same RWA-strip with $n_{\Sigma}=n+k$ excitations~\cite{MostafaThesis}, where $n$ indicates the number of photons in the readout resonator (an RWA-strip with $n_{\Sigma}$ excitations is defined as the subspace of states $\ket{k,n}$ such that $n + k = n_\Sigma$).

The non-RWA terms in the coupling are given by 
\begin{align}
  \label{SM-H_non_RWA}
  H_{\rm non-RWA}
  = g\lambda \left[ \frac{3}{2} (b^\dagger b)b^\dagger a^\dagger - \frac{1}{2}{b^\dagger}^3a + \frac{1}{4}{b^\dagger}^3a^\dagger \right] + {\rm H.c.}
\end{align}
There are two types of non-RWA terms in $H_{\rm non-RWA}$.
The first type of non-RWA terms include the terms $(b^\dagger b) b^\dagger a^\dagger$ and ${b^\dagger}^3a$ (as well as their Hermitian-conjugated terms) that couple states of two RWA-strips separated by 2 excitations.
The second type of non-RWA terms are ${b^\dagger}^3a^\dagger$ and $b^3a$ that couple RWA-strips separated by 4 excitations~\cite{MostafaThesis}.
These non-RWA couplings open the way to measurement-induced transitions (MIST) between, e.g., the qubit ground state to high-energy qubit eigenstates during qubit readout.
It is not possible to find a suitable operation point $(g_c,g_l)$ that eliminates all non-RWA terms.

More generally, we investigate the matrix elements involved in moving the system from one RWA strip to another one with two more excitations.
The matrix elements of $\bar{H}_{\rm coupling}$  involved in the transition $\ket{k, n} \rightarrow \ket{k + 1, n + 1}$ are
\begin{align}
  \bbraket{k + 1, n + 1}{\bar \Phi_{\rm q} a^\dagger}{k, n}
    &= \sqrt{n + 1} \left( \sqrt{k + 1} + \frac{3 \lambda}{2} \left( k + 1 \right)^{3/2} \right) \\
  \bbraket{k + 1, n + 1}{i \bar{Q}_{\rm q} a^\dagger}{k, n}
    &= \sqrt{n + 1} \left( -\sqrt{k + 1} + \frac{3 \lambda}{2} \left( k + 1 \right)^{3/2} \right)
\end{align}
and the matrix elements in involved in the transition $\ket{k, n} \rightarrow \ket{k + 3, n - 1}$ are
\begin{align}
  \bbraket{k + 3, n - 1}{\bar{\Phi}_{\rm q} a}{k, n}
    &= - \sqrt{n} \frac{\lambda}{4} \sqrt{(k + 1) (k + 2) (k + 3)} \\
  \bbraket{k + 3, n - 1}{i \bar{Q}_{\rm q} a}{k, n}
    &= \sqrt{n} \frac{3 \lambda}{4} \sqrt{(k + 1) (k + 2) (k + 3)}
  \, .
\end{align}

\end{document}